\documentclass[prd,amsmath,amssymb,longbibliography,superscriptaddress,twocolumn,nofootinbib,10pt,floatfix]{revtex4}
\pdfoutput=1

\usepackage{graphicx}
\usepackage{dcolumn}
\usepackage{bm}
\usepackage{amssymb}
\usepackage{latexsym}
\usepackage{booktabs}
\usepackage{amsmath}
\usepackage{multirow}
\usepackage[colorlinks=true, linkcolor=blue, citecolor=blue]{hyperref}

\begin{document}

\title{Finite-Core Signatures in LISA-Band Wave-Optics Lensing by Low-Mass Dark Matter Halos}

\author{Dejiang Li}
\affiliation{Department of Physics, and Collaborative Innovation Center for Quantum Effects and Applications, Hunan Normal University, Changsha 410081, China;}
\affiliation{College of Information Science and Engineering, Hunan Normal University, Changsha, Hunan 410081, People's Republic of China;}

\author{Tonghua Liu}
\email{liutongh@yangtzeu.edu.cn}
\affiliation{School of Physics and Optoelectronic Engineering, Yangtze University, Jingzhou, 434023, China;}

\author{Kai Liao}
\affiliation{School of Physics and Technology, Wuhan University, Wuhan 430072, China;}

\author{Beining Xia}
\affiliation{Department of Physics, and Collaborative Innovation Center for Quantum Effects and Applications, Hunan Normal University, Changsha 410081, China;}
\affiliation{College of Information Science and Engineering, Hunan Normal University, Changsha, Hunan 410081, People's Republic of China;}

\author{Cuihong Wen$^{\ast}$}
\email{cuihongwen@hunnu.edu.cn}
\affiliation{Department of Physics, and Collaborative Innovation Center for Quantum Effects and Applications, Hunan Normal University, Changsha 410081, China;}
\affiliation{College of Information Science and Engineering, Hunan Normal University, Changsha, Hunan 410081, People's Republic of China;}

\author{Jieci Wang$^{\ddagger}$}
\email{jcwang@hunnu.edu.cn}
\affiliation{Department of Physics, and Collaborative Innovation Center for Quantum Effects and Applications, Hunan Normal University, Changsha 410081, China;}
\affiliation{Hunan Research Center of the Basic Discipline for Quantum Effects and Quantum Technologies, Hunan Normal University, Changsha, Hunan 410081, China;}

\begin{abstract}
LISA-band gravitational waves from massive binary black holes can be diffractively lensed by low-mass dark matter halos and subhalos, so their frequency-dependent amplification can probe the inner density profile.  We isolate the generic finite-core part of this signal by comparing fixed-mass Navarro-Frenk-White (NFW) and cored-NFW lenses and propagating both profiles to the complex wave-optics amplification factor.  A finite core smooths the time-delay response and reshapes the diffraction peak; an NFW template with a lower concentration can mimic part of the effect, but structured complex residuals remain after time and phase alignment.  The residual peaks for intermediate cores, $r_c/r_s\simeq0.25$--$0.3$.  An SIDM-inspired isothermal-core profile gives the same qualitative response, showing that the signal is not an artifact of one cored parameterization.  For a fiducial LISA source, an appreciable mismatch requires favorable near alignment and $M_{\rm vir}\gtrsim 10^7M_\odot$.  The result is a finite-core baseline for isolated line-of-sight halos and for subhalos perturbing strongly lensed macro-images.
\end{abstract}

\maketitle

\section{Introduction}
\label{sec:introduction}

A central question in dark-matter physics is whether low-mass halos have cuspy or cored inner density profiles.  Collisionless cold dark matter predicts approximately NFW-like cusps in numerical simulations~\cite{1997ApJ...490..493N}.  Finite-density cores are nevertheless suggested by the long-standing cusp-core problem in dwarf and low-surface-brightness galaxies~\cite{1995ApJ...447L..25B,2011AJ....142...24O,2015PNAS..11212249W,2017ARA&A..55..343B}. Such cores are not tied to a single microscopic explanation. Self-interacting dark matter (SIDM) can thermalize the inner halo and form an approximately isothermal region~\cite{2000PhRvL..84.3760S,2013MNRAS.430...81R,2018PhR...730....1T,2025RvMP...97d5004A}; Fuzzy dark matter (FDM) predicts solitonic cores inside extended halos~\cite{2000PhRvL..85.1158H,2014NatPh..10..496S,2017PhRvD..95d3541H,2016PhR...643....1M,2020PrPNP.11303787N}; and baryonic feedback can also flatten the central profile in some systems~\cite{2011AJ....142...24O,2015PNAS..11212249W}.  A useful gravitational-wave test should therefore make a clear distinction between a generic finite-core imprint and a model-specific FDM or SIDM interpretation.

Gravitational waves provide a particularly clean way to probe projected mass because their phase is coherent.  When the wavelength is comparable to the lensing time delay, lensing is not described only by image positions and magnifications.  Instead, the observed waveform is multiplied by a complex amplification factor whose amplitude and phase vary with frequency~\cite{1974IJTP....9..425O,1999PThPS.133..137N,2003ApJ...595.1039T}.  This is the wave-optics regime.  We focus on the Laser Interferometer Space Antenna (LISA), whose massive-binary-black-hole (MBHB) sources radiate mainly in the $10^{-4}$--$1\,{\rm Hz}$ band~\cite{2023LRR....26....2A}.  The relevant dimensionless frequency is $w=8\pi G M_{\rm Lz}f/c^3$, where $M_{\rm Lz}$ is the redshifted lens mass scale and $f$ is the observed GW frequency.  Thus $w\sim1$ at $M_{\rm Lz}\simeq 8.1\times10^6M_\odot \left({10^{-3}{\rm Hz}}/{f}\right)$.
This simple scaling fixes the physical mass range of the problem.  In the LISA band, diffractive lensing is naturally produced by low-mass halos and subhalos, roughly $10^5$--$10^9M_\odot$ depending on the frequency range with the largest signal-to-noise.  A $10^{11}M_\odot$ galaxy-scale halo is instead in the high-$w$ geometric-optics regime for most LISA frequencies.  Such a galaxy can act as a macro lens, but the diffractive feature studied here should come from a smaller halo or subhalo.

This leads to two concrete astrophysical settings.  First, an isolated low-mass halo along the line of sight can weakly lens a single observed MBHB waveform.  Second, a galaxy-scale strong lens can create macro-images, while a much smaller dark-matter subhalo near one macro-image adds a local frequency-dependent perturbation.  The second case is especially important because macro magnification can raise the signal-to-noise ratio of the perturbed image.  Recent work has developed LISA wave-optics forecasts for low-mass halos, dark substructures, compact dark matter, and FDM solitonic cores~\cite{Gao2022LensedSMBHBLISA,Caliskan2023LISAObservability,Caliskan2023LensingDMHalos,2023PhRvD.108j3529T,2022PhRvD.106b3018G,2025PhRvD.111b4068B,2025JCAP...07..025S,Liu2025MicrolensingDMML}.  Related strong-lensing studies, have addressed the identification of strongly lensed GW events, host-galaxy priors, mock catalogs, subhalo lensing, and overlapping LISA images~\cite{2026ApJS..283...31L,2026PhRvD.113h3009L,2026ApJS..284...56L,Sun2026LensedGWMockSpace,Wu2026LISABeatPatterns}.  These studies motivate a local calculation that can later be embedded into a full strong-lens population.

The purpose of this paper is deliberately narrower than a complete dark-matter model test.  We ask what changes in the wave-optics signal when an NFW cusp is replaced by a finite-density core at fixed virial mass.  This question is useful because FDM, SIDM, baryonic feedback, and phenomenological cored profiles can all alter the inner potential.  Before interpreting a future signal as evidence for one microscopic model, one should know the generic finite-core baseline that any such model would have to exceed.

We construct this baseline with a one-parameter cored-NFW profile and compare it directly with a cuspy NFW halo.  The advantage of this setup is that the total virial mass is held fixed, so the comparison isolates the redistribution of mass in the inner halo rather than a trivial change in total mass.  We then propagate both profiles through the same chain: three-dimensional density, projected convergence, lensing potential, time-delay response, complex wave-optics amplification, and finally a detector-weighted LISA mismatch estimate.  We also include an SIDM-inspired isothermal-core NFW cross-check to verify that the qualitative response is not peculiar to the chosen cored-NFW formula.


The paper is organized as follows.  Section~\ref{sec:formalism_models} defines the astrophysical setup, wave-optics quantities, and halo profiles.  Section~\ref{sec:wave_optics_signatures} shows how finite cores modify $\kappa$, $\psi$, $I(\tau)$, and $F(w)$.  Section~\ref{sec:distinguishability} tests whether NFW templates can absorb the cored signal.  Section~\ref{sec:detectability} gives a detector-weighted LISA estimate.  Section~\ref{sec:conclusions} summarizes the physical interpretation and limitations.

\section{Physical setup and halo models}
\label{sec:formalism_models}

\subsection{Astrophysical interpretation of the lens}
\label{subsec:astrophysical_setup}

The lens is a low-mass dark matter halo or subhalo in the LISA wave-optics mass range.  We denote the lens and source redshifts by $z_{\rm L}$ and $z_{\rm S}$, and the corresponding angular-diameter distances by $D_{\rm L}$, $D_{\rm S}$, and $D_{\rm LS}$.  Unless stated otherwise, the numerical examples use $z_{\rm L}=0.5$ and $z_{\rm S}=2.0$.  The halo is labeled by its virial mass $M_{\rm vir}$, virial radius $R_{\rm vir}$, scale radius $r_s$, and concentration $c_{\rm vir}=R_{\rm vir}/r_s$.  When we convert from dimensionless frequency to the observed GW frequency, we use the redshifted lens mass scale $M_{\rm Lz}\simeq(1+z_{\rm L})M_{\rm vir}$.

The same local calculation has two possible physical readings.  In the isolated-halo case, the low-mass halo lies along the line of sight and produces a weak frequency-dependent amplification of one MBHB waveform.  In the strong-lens case, a galaxy-scale lens first produces macro-images in the geometric-optics regime, and the low-mass subhalo near one macro-image adds the local wave-optics perturbation studied here.  The present paper models only this local perturber.  It does not include external convergence, external shear, macro-image parity, or a cosmological subhalo population.

The response-level figures use $M_{\rm vir}=10^6,10^7,10^8\,M_\odot$ and dimensionless impact parameters $y=3,5,8$, where $y$ is the separation between the source and the lens center in the units used by the diffraction integral.  These relatively large values show the morphology of the transfer function in a weak single-image regime.  The detector-weighted scan in Sec.~\ref{sec:detectability} extends to $10^9M_\odot$ but not beyond it, because more massive galaxy-scale lenses are already mostly geometric-optics-like across the LISA frequencies that carry most of the signal-to-noise.  The smaller impact parameters used in that scan reflect a simple fact: for the fiducial LISA source adopted here, an observable distortion requires close alignment.

The source is a LISA MBHB signal.  For a total binary mass $M_{\rm BBH}\sim10^5$--$10^7M_\odot$, the inspiral--merger--ringdown power lies in the mHz band.  This is the band in which diffraction by $10^5$--$10^9M_\odot$ halos can modify both the waveform amplitude and phase.  This mass-frequency connection is why the present analysis is framed around LISA rather than pulsar-timing arrays or ground-based detectors.

\subsection{Wave-optics quantities}
\label{subsec:wave_optics_formalism}

We work with dimensionless lens-plane and source-plane coordinates, $\mathbf{x}$ and $\mathbf{y}$.  The physical lens-plane radius is denoted by $\xi$, while $y\equiv|\mathbf{y}|$ is the dimensionless impact parameter.  For a projected surface density $\Sigma(\xi)$, the convergence is $\kappa=\Sigma/\Sigma_{\rm cr}$, where
$\Sigma_{\rm cr}=c^2D_{\rm S}/(4\pi G D_{\rm L}D_{\rm LS})$ in ordinary units.  The dimensionless lensing potential $\psi$ satisfies $\nabla_x^2\psi=2\kappa$.  The Fermat potential is
\begin{equation}
\phi(\mathbf{x},\mathbf{y})=\frac{1}{2}|\mathbf{x}-\mathbf{y}|^2-\psi(\mathbf{x})-\phi_{\rm m}(\mathbf{y}),
\label{eq:fermat}
\end{equation}
where $\phi_{\rm m}$ is chosen so that the minimum of $\phi$ is zero.  This subtraction fixes an irrelevant overall time delay and makes the response easier to compare between profiles.

The lensed and unlensed frequency-domain waveforms are related by $\tilde h_{\rm L}(f)=F(f)\tilde h_0(f)$.  In dimensionless variables the wave-optics amplification factor is
\begin{equation}
F(w)=\frac{w}{2\pi i}\int d^2x\,\exp[iw\phi(\mathbf{x},\mathbf{y})],
\ w=8\pi G M_{\rm Lz} f/c^3 .
\label{eq:diffraction_integral}
\end{equation}
Here $f$ is the observed GW frequency and $M_{\rm Lz}$ is the redshifted lens mass scale.  The condition $w\sim1$ identifies the diffractive transition.  This is the reason why the calculations below focus on $10^6$--$10^9M_\odot$ perturbers rather than on $10^{11}M_\odot$ galaxy halos.

It is often useful to view the same response in the time-delay domain.  We define
\begin{equation}
I(\tau)=\int d^2x\,\delta[\phi(\mathbf{x},\mathbf{y})-\tau],\
F(w)=\frac{w}{2\pi i}\int d\tau\,I(\tau)e^{iw\tau} .
\label{eq:Itau_def}
\end{equation}
The dimensionless variable $\tau$ is the time delay associated with $\phi$.  Physically, $I(\tau)$ measures how much lens-plane area lies on a constant-Fermat-potential contour.  A cusp produces a sharper response, while a finite core smooths the central part of this distribution and therefore changes the oscillatory structure of $F(w)$.

\subsection{NFW and cored-NFW baseline}
\label{subsec:nfw_cored_nfw}

The cuspy benchmark is the NFW profile~\cite{1997ApJ...490..493N},
\begin{equation}
\rho_{\rm NFW}(r)=\frac{\rho_s}{(r/r_s)(1+r/r_s)^2} .
\label{eq:nfw_density}
\end{equation}
Here $r$ is the three-dimensional radius, $r_s$ is the scale radius, and $\rho_s$ is fixed by the virial mass.  For concentration $c_{\rm vir}=R_{\rm vir}/r_s$, the mass normalization is
\begin{equation}
M_{\rm vir}=4\pi\rho_s r_s^3\left[\ln(1+c_{\rm vir})-\frac{c_{\rm vir}}{1+c_{\rm vir}}\right].
\label{eq:nfw_mass_norm}
\end{equation}

To isolate the effect of replacing the cusp by a finite-density core, we use the cored-NFW form adopted in strong-lensing studies~\cite{2022MNRAS.510...54A},
\begin{equation}
\rho_{\rm cNFW}(r;q)=\frac{\rho_0(q)}{(r/r_s+q)(1+r/r_s)^2},
\qquad q\equiv\frac{r_c}{r_s} .
\label{eq:cnfw_density}
\end{equation}
The parameter $r_c$ is the core radius, and $q$ is the core size in units of $r_s$.  For $q>0$ the central density is finite; the NFW form is recovered continuously as $q\rightarrow0$.  We keep $M_{\rm vir}$ fixed when changing $q$, so the density normalization $\rho_0(q)$ is determined by
\begin{equation}
M_{\rm vir}=4\pi\rho_0(q)r_s^3\int_0^{c_{\rm vir}}\frac{u^2\,du}{(u+q)(1+u)^2},
\label{eq:cnfw_mass_norm}
\end{equation}
where $u=r/r_s$.  This fixed-mass prescription is important: it changes the inner mass distribution without changing the total mass inside $R_{\rm vir}$.  The resulting difference in $F(w)$ can therefore be traced to the inner profile rather than to an overall mass rescaling.

For both profiles, we project the three-dimensional density to obtain $\kappa$, solve for $\psi$, compute $I(\tau)$, and finally obtain $F(w)$.  The NFW convergence and potential are checked against standard analytic expressions~\cite{1996A&A...313..697B,2000ApJ...534...34W}.  Numerical projection details are given in Appendix~\ref{app:numerical_definitions}.

\subsection{SIDM-inspired cross-check}
\label{subsec:sidm_profile}

The cored-NFW profile is a phenomenological baseline, not a unique prediction of SIDM, FDM, or baryonic feedback.  To verify that the result is not an artifact of this particular one-parameter form, we also use an SIDM-inspired isothermal-core NFW (IC-NFW) profile following the implementation adopted in Ref.~\cite{2022PhRvD.106b3018G}.  The profile is schematically
\begin{equation}
\rho_{\rm ICNFW}(r)=
\begin{cases}
\rho_{\rm iso}(r), & r<r_1,\\
\rho_{\rm NFW}(r), & r\ge r_1,
\end{cases}
\label{eq:icnfw_profile}
\end{equation}
where the inner self-gravitating isothermal solution is matched to the outer NFW halo at $r_1=s r_s$ by imposing continuity of the density and enclosed mass.  The parameter $s$ controls the radial extent of the isothermal core.  In this work we use $s=0.1$ and $s=1$ only as representative cross-checks; we do not infer a self-interaction cross section.  The matching equations are given in Appendix~\ref{app:icnfw_details}.

\section{Response-level wave-optics signatures}
\label{sec:wave_optics_signatures}

We first study the transfer function itself, before applying a detector noise curve.  The goal of this section is not to claim that every curve shown here is directly detectable.  Instead, we identify how a finite core reshapes the lensing potential and how that reshaping propagates into $I(\tau)$ and $F(w)$.  The detector-weighted relevance of the same effect is addressed separately in Sec.~\ref{sec:detectability}.

Unless otherwise stated, the cored-NFW curves are shown for $r_c/r_s=0.1,0.3,0.5$ and are compared with the corresponding NFW halo at the same virial mass.  The three columns in Figs.~\ref{fig:kappa_cnfw}--\ref{fig:Fw_cnfw} correspond to $M_{\rm vir}=10^6,10^7,10^8\,M_\odot$.  In the wave-optics panels we use $y=3,5,8$ to illustrate the weak single-image response at moderate source-lens offsets.  These values are useful for displaying the morphology of the finite-core distortion; the detectable region for the fiducial LISA waveform lies at much smaller $y$ in Sec.~\ref{sec:detectability}.

For each comparison, the NFW and cored-NFW halos are assigned the same concentration at fixed $M_{\rm vir}$.  When a mass-dependent concentration is needed, we use an extrapolated concentration--mass relation motivated by Duffy et al.~\cite{2008MNRAS.390L..64D}.  The fixed-mass, fixed-concentration comparison ensures that the differences below are driven by the central core.

\subsection{Convergence and lensing potential}
\label{subsec:results_kappa_psi}

\begin{figure*}[t]
    \centering
    \includegraphics[width=\textwidth]{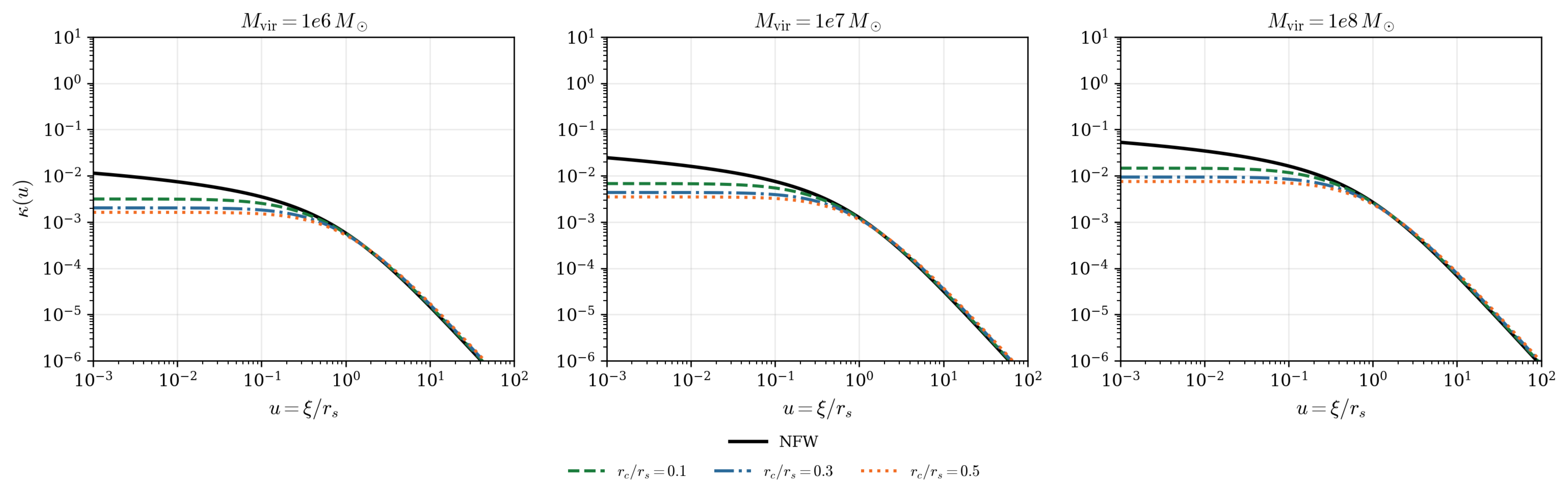}
    \caption{
    Convergence \(\kappa(u)\) for NFW and cored-NFW halos, where \(u=\xi/r_s\).  The three panels correspond to \(M_{\rm vir}=10^6,10^7,10^8\,M_\odot\).  The black solid curve denotes the NFW profile, while the dashed, dash-dotted, and dotted curves correspond to cored-NFW profiles with \(r_c/r_s=0.1,0.3,0.5\), respectively.  A finite core suppresses the central convergence, with stronger suppression for larger \(r_c/r_s\).
    }
    \label{fig:kappa_cnfw}
\end{figure*}

Figure~\ref{fig:kappa_cnfw} shows the convergence \(\kappa(u)\), with \(u=\xi/r_s\).  A finite core suppresses the central convergence and flattens the inner projected density relative to NFW, with stronger suppression for larger \(r_c/r_s\).

At \(u\gtrsim1\), the profiles approach one another because the core mainly modifies the inner halo while the fixed-mass normalization preserves the total mass inside \(R_{\rm vir}\).  Thus the cored profiles are not obtained by a uniform lowering of the NFW density; rather, the central cusp is regularized while the outer halo remains close to NFW.

\begin{figure*}[t]
    \centering
    \includegraphics[width=\textwidth]{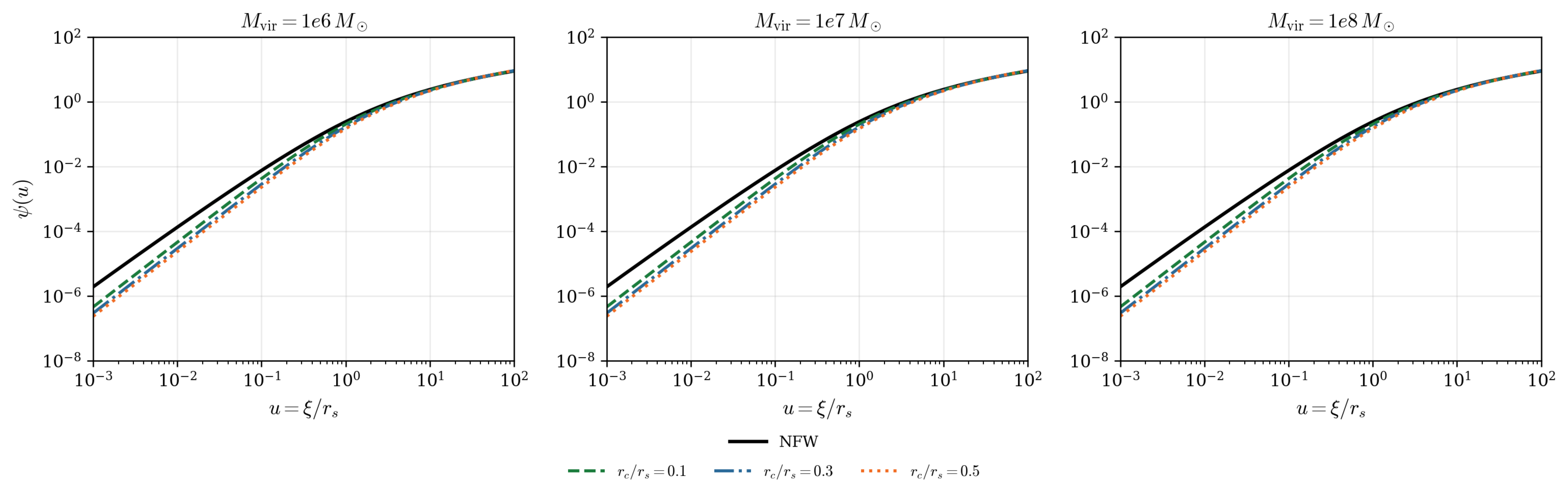}
    \caption{
    Lensing potential \(\psi(u)\) for the same halo models as in Fig.~\ref{fig:kappa_cnfw}.  The cored-NFW profiles produce a shallower inner potential than the NFW halo, and the difference increases with the core size.  Since \(\psi\) enters the Fermat potential, these inner-potential differences are expected to propagate into both the time-domain and frequency-domain wave-optics observables.
    }
    \label{fig:psi_cnfw}
\end{figure*}

The corresponding lensing potential is shown in Fig.~\ref{fig:psi_cnfw}.  With the same potential convention, the cored-NFW profiles produce a shallower inner potential than NFW, with a difference that increases with \(r_c/r_s\).  Since \(\psi\) enters the Fermat potential in Eq.~\eqref{eq:fermat}, this localized change in the inner mass distribution modifies the constant-\(\phi\) contours that determine \(I(\tau)\) and hence the oscillatory structure of \(F(w)\).

\subsection{Time-domain diffraction integral \(I(\tau)\)}
\label{subsec:results_Itau}

\begin{figure*}[t]
    \centering
    \includegraphics[width=\textwidth]{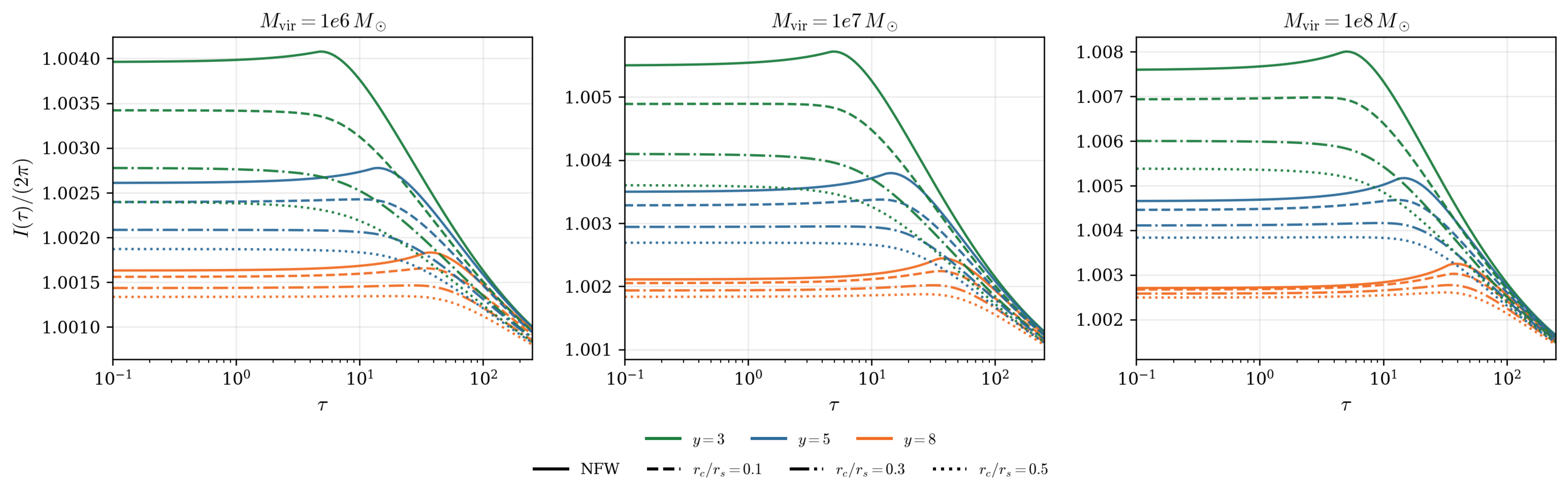}
    \caption{
    Time-domain diffraction integral \(I(\tau)/(2\pi)\) for NFW and cored-NFW halos.  The three columns correspond to \(M_{\rm vir}=10^6,10^7,10^8\,M_\odot\).  The colors denote the impact parameters \(y=3,5,8\), and the different line styles denote NFW and cored-NFW profiles with \(r_c/r_s=0.1,0.3,0.5\).  A finite core lowers and broadens the time-domain response, especially at small impact parameter.
    }
    \label{fig:Itau_cnfw}
\end{figure*}

Figure~\ref{fig:Itau_cnfw} shows the time-domain diffraction integral \(I(\tau)/(2\pi)\).  For each halo mass, the colored curves correspond to different impact parameters, while the line styles distinguish NFW from cored-NFW profiles.

The NFW curves generally exhibit a more concentrated response in \(\tau\), especially for small impact parameter.  For \(y=3\), the cusp enhances the contribution from wavefronts probing the central lens, producing a more pronounced time-domain feature.

Replacing the central cusp by a finite core smooths this response.  At fixed \(M_{\rm vir}\) and \(y\), increasing \(r_c/r_s\) lowers the height of \(I(\tau)/(2\pi)\) and broadens the feature in \(\tau\).  The arrival-time distribution therefore becomes less sharply concentrated, reflecting the shallower inner potential shown in Fig.~\ref{fig:psi_cnfw}.

The core-induced modification is strongest at small impact parameter and increases with halo mass.  A smaller \(y\) allows the wavefront to probe the inner potential more efficiently, while a larger \(M_{\rm vir}\) produces a stronger lensing perturbation.  Across the three masses shown, the qualitative effect remains the same: the central response is suppressed and smoothed relative to the NFW case.

\subsection{Frequency-domain amplification factor \(F(w)\)}
\label{subsec:results_Fw}

\begin{figure*}[t]
    \centering
    \includegraphics[width=\textwidth]{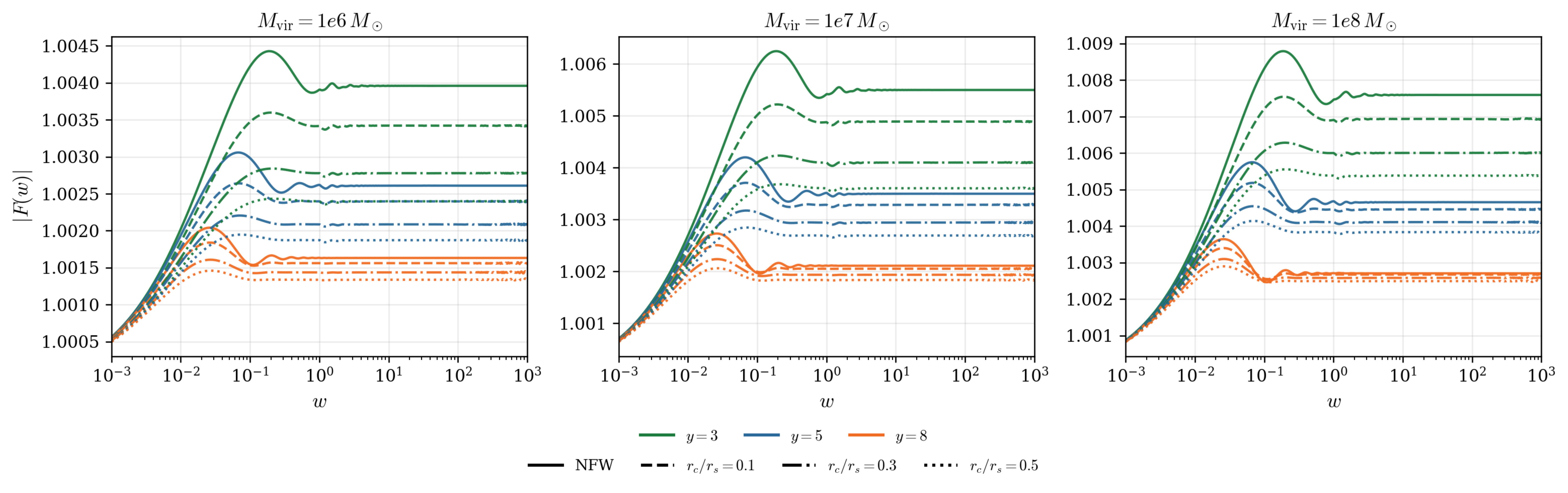}
    \caption{
    Frequency-domain amplification factor \(|F(w)|\) for NFW and cored-NFW halos.  The notation is the same as in Fig.~\ref{fig:Itau_cnfw}.  The finite core changes both the first diffraction peak and the subsequent oscillatory structure.  The effect is strongest for small impact parameter and becomes more pronounced for larger core size.
    }
    \label{fig:Fw_cnfw}
\end{figure*}

Figure~\ref{fig:Fw_cnfw} shows the corresponding frequency-domain amplification factor \(|F(w)|\) over the dimensionless frequency range relevant for the transition from weak diffraction to the high-frequency limit.

The frequency-domain behavior is the Fourier counterpart of the time-domain response.  A more localized \(I(\tau)\) produces stronger oscillatory structure in \(F(w)\), whereas the core-induced smoothing of \(I(\tau)\) reduces these oscillations.

As shown in Fig.~\ref{fig:Fw_cnfw}, the NFW halo generally produces a sharper first diffraction peak and more visible post-peak oscillations, especially for \(y=3\).  Increasing \(r_c/r_s\) lowers and smooths the first peak and suppresses the residual oscillatory structure.  This is not a simple global rescaling of \(|F(w)|\), but a frequency-dependent distortion caused by the core-induced change in the Fermat-potential contours.

The dependence on impact parameter follows the same pattern as in the time domain.  The effect weakens as \(y\) increases because the wavefront probes the central density profile less efficiently.  Nevertheless, Fig.~\ref{fig:Fw_cnfw} shows that a finite core leaves a coherent imprint on the shape of \(|F(w)|\), motivating the full complex comparison in Sec.~\ref{sec:distinguishability}.

\subsection{SIDM-inspired comparison}
\label{subsec:results_sidm}

\begin{figure*}[t]
    \centering
    \includegraphics[width=\textwidth]{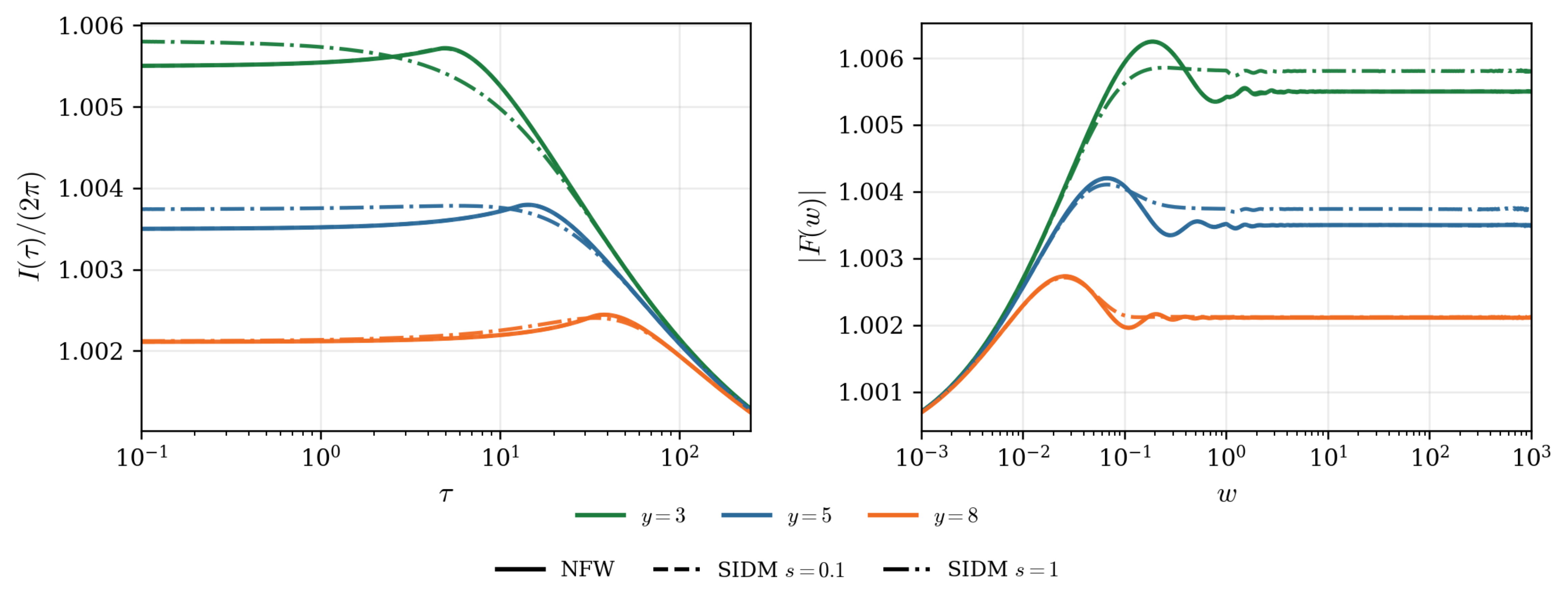}
    \caption{
    SIDM-inspired IC-NFW comparison for \(M_{\rm vir}=10^7\,M_\odot\), \(z_{\rm L}=0.5\), and \(z_{\rm S}=2.0\).  Left: time-domain diffraction integral \(I(\tau)/(2\pi)\).  Right: frequency-domain amplification factor \(|F(w)|\).  The solid curves denote the NFW halo, while the dashed and dash-dotted curves denote IC-NFW profiles with \(s=0.1\) and \(s=1\), respectively.  The colors correspond to \(y=3,5,8\).  The SIDM-inspired core modifies both \(I(\tau)\) and \(|F(w)|\).  The \(s=0.1\) curves are close to NFW, while the \(s=1\) case shows a more visible deviation, especially at smaller impact parameter.
    }
    \label{fig:sidm_comparison}
\end{figure*}

We now check whether the qualitative behavior found for the phenomenological cored-NFW profile also appears in a more physically motivated cored halo.  SIDM simulations and gravothermal analyses have shown that self-interactions can generate central cores and approximately isothermal inner regions~\cite{2012MNRAS.423.3740V,2011MNRAS.415.1125K,2013MNRAS.430..105P,2016PhRvL.116d1302K,2013MNRAS.430...81R}.  Figure~\ref{fig:sidm_comparison} compares the NFW halo with the SIDM-inspired IC-NFW profiles described in Sec.~\ref{subsec:sidm_profile}.  We show representative cases with matching radii \(s=0.1\) and \(s=1\).

The SIDM-inspired profiles produce the same broad class of effects as the cored-NFW model: the time-domain response is reshaped, and the frequency-domain amplification factor shows a modified diffraction peak and altered post-peak structure.  The effect is small for \(s=0.1\), where the isothermal core is matched to the NFW profile at a relatively small radius.  For \(s=1\), the inner isothermal region extends to \(r_s\), and the deviation from NFW becomes more visible, especially for \(y=3\) and, to a lesser extent, \(y=5\).  The \(y=8\) curves remain close to NFW because the wavefront probes the central region less efficiently.

There are also quantitative differences between the IC-NFW and cored-NFW cases.  In the IC-NFW construction, the inner isothermal core is matched to the outer NFW halo by imposing continuity of both density and enclosed mass at \(r_1=s r_s\).  This matching prescription redistributes the inner mass differently from the simple cored-NFW profile.  As a result, the detailed changes in \(I(\tau)\) and \(|F(w)|\) are not identical to those in Figs.~\ref{fig:Itau_cnfw} and~\ref{fig:Fw_cnfw}.  In particular, depending on \(s\) and \(y\), the IC-NFW profile can shift the time-domain response and high-frequency level in a way that is not identical to the phenomenological cored-NFW result.

The important point, however, is that the qualitative conclusion is unchanged.  Replacing the NFW cusp by a finite-density, physically motivated core leaves a coherent wave-optics imprint.  In Fig.~\ref{fig:sidm_comparison}, this imprint is clearest for the larger matching radius and smaller impact parameters.  This supports the interpretation that the wave-optics signatures identified above are not artifacts of a particular analytic cored-NFW parameterization.  Instead, they reflect a more general sensitivity of diffractive gravitational-wave lensing to the inner structure of dark matter halos.

\section{Distinguishability from NFW templates}
\label{sec:distinguishability}

The results in Sec.~\ref{sec:wave_optics_signatures} show that a finite central core modifies both the time-domain response and the frequency-domain amplification factor.  We now ask a more quantitative question: can these core-induced wave-optics signatures be mimicked by an NFW halo with different lens parameters?  This is an important test because, in an actual observation, the lens parameters are not known a priori and must be inferred jointly with the source and waveform parameters~\cite{2015PhRvD..91d2003V,2026ApJS..283...31L,2023MNRAS.526..682G}.  A difference between cored-NFW and NFW halos is physically meaningful only if it cannot be completely absorbed by reasonable variations of the NFW template.

In this section we treat the cored-NFW signal as the fiducial signal and fit it with a family of NFW templates.  For the two-dimensional template scan, we use a fiducial cored-NFW configuration with \(M_{\rm vir}=10^7\,M_\odot\), \(z_{\rm L}=0.5\), \(z_{\rm S}=2.0\), \(c_{\rm true}=10\), \(y_{\rm true}=3\), and \(r_c/r_s=0.1\).  The NFW templates are generated at the same virial mass and redshifts, but with varying impact parameter \(y_{\rm NFW}\) and concentration \(c_{\rm NFW}\).

\subsection{NFW-template fitting setup}
\label{subsec:template_setup}

We choose \(y_{\rm NFW}\) and \(c_{\rm NFW}\) as the adjustable NFW-template parameters because they represent the two most relevant degeneracy directions for the present problem.  The impact parameter \(y\) is an extrinsic geometrical parameter describing the source-lens alignment.  It strongly affects the overall strength of the lensing perturbation, the location of the time-domain feature in \(I(\tau)\), and the height and position of the diffraction peak in \(F(w)\).  Since the true source position relative to the lens is generally unknown, any realistic comparison between cored and cuspy halos must allow \(y\) to vary.

The concentration \(c\), on the other hand, is the main shape parameter of an NFW halo at fixed virial mass and redshift.  It controls how centrally concentrated the projected mass distribution is, and therefore directly affects the lensing potential and the wave-optics response.  Moreover, halo concentrations have intrinsic scatter, and their extrapolation to the low-mass regime relevant for wave-optics lensing is uncertain~\cite{2008MNRAS.390L..64D,2014MNRAS.441.3359D,2016MNRAS.460.1214L}.  Allowing \(c_{\rm NFW}\) to vary therefore gives the NFW template a conservative freedom to mimic the shallower central potential of a cored halo.

We keep \(M_{\rm vir}\), \(z_{\rm L}\), and \(z_{\rm S}\) fixed in this scan.  This choice is deliberate: the goal is not to perform a full parameter-estimation analysis, but to test whether the core-induced distortion can be absorbed by the most relevant NFW nuisance parameters while remaining in the same halo-mass framework.  Varying the halo mass would also change the characteristic lensing scale and the mapping between physical frequency and dimensionless frequency, introducing an additional degeneracy that is not the focus of this section.  The present scan therefore isolates the degeneracy between a finite core, source-lens alignment, and NFW concentration.

We compare the full complex frequency-domain amplification factors rather than only their magnitudes.  This is essential because the core imprint is carried by both the envelope $|F(w)|$ and the phase of the complex response.  To connect the comparison with GW data analysis, we use the standard maximized mismatch between two frequency-domain waveforms,
\begin{equation}
{\cal M}_{12}=1-
\max_{M_{\rm vir},y}
\frac{\langle h_1|h_2\rangle}
{\sqrt{\langle h_1|h_1\rangle\langle h_2|h_2\rangle}} .
\label{eq:standard_mismatch}
\end{equation}
The inner product is the usual noise-weighted one, $\langle a|b\rangle=4\,\mathrm{Re}\int df\,\tilde a(f)\tilde b^*(f)/S_n(f)$, with $S_n(f)$ the one-sided detector noise power spectral density.  At the transfer-function level, before choosing a detector noise curve, we apply the same idea to $F(w)$ with a logarithmic weight in $w$.  The unnormalized residual norm plotted in Figs.~\ref{fig:Dfull_heatmap} and~\ref{fig:Dmin_coretrend} is defined in Appendix~\ref{app:profile_residual}; it is used only to display the NFW degeneracy valley.  Detector-level statements in Sec.~\ref{sec:detectability} use Eq.~\eqref{eq:standard_mismatch} directly.

In the template scan, the NFW impact parameter and concentration are varied over \(0.2\le y_{\rm NFW}\le 10\) and \(5\le c_{\rm NFW}\le 30\), respectively.  The numerical response-level scan is performed over \(10^{-3}\le w\le10^2\), which resolves both the diffraction onset and the high-frequency approach to geometric optics.

\subsection{Heat maps in the \(y-c\) plane}
\label{subsec:heatmap_yc}

\begin{figure}[t]
    \centering
    \includegraphics[width=\columnwidth]{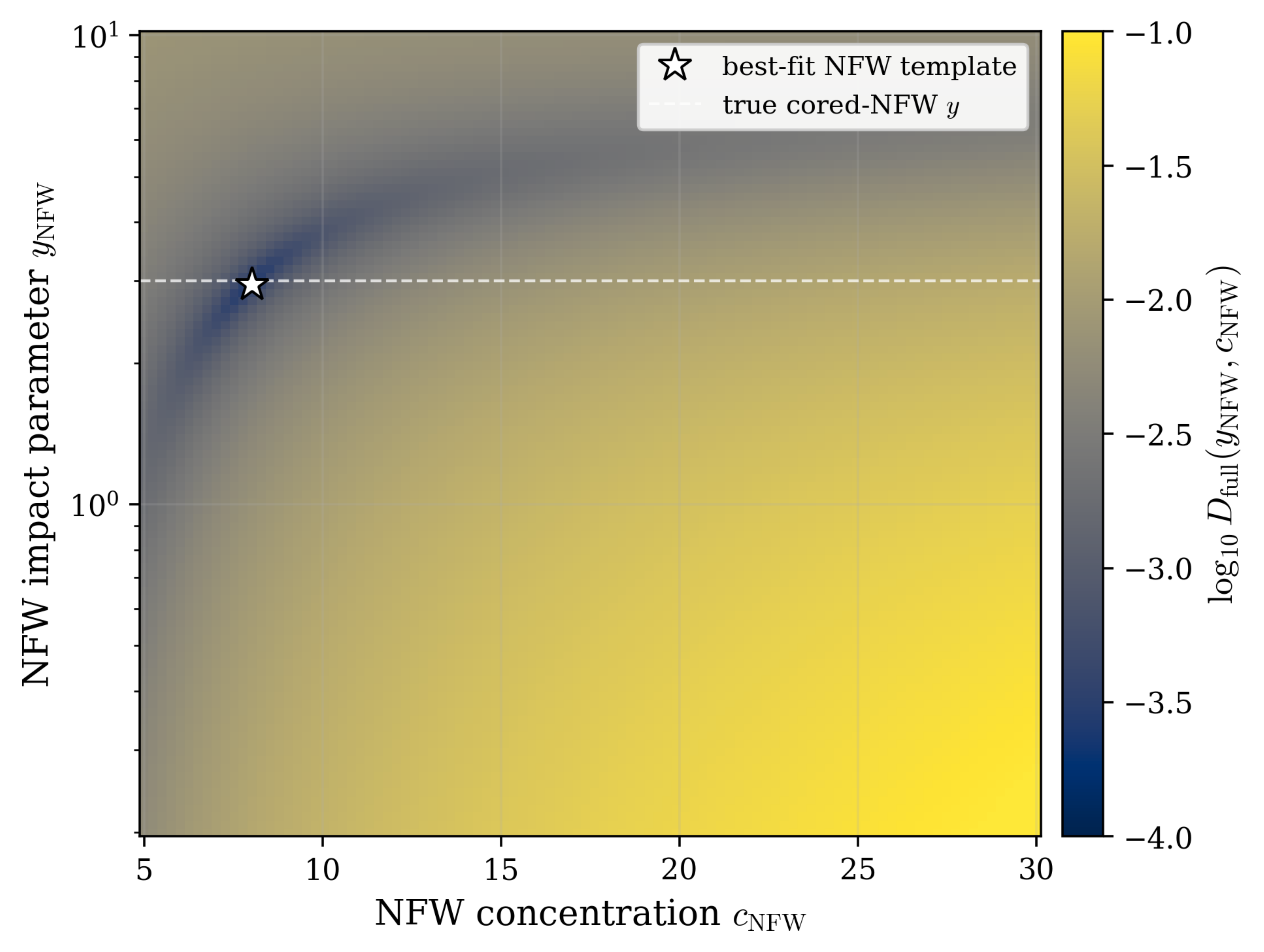}
    \caption{
    Template-fitting residual in the \(y_{\rm NFW}\)-\(c_{\rm NFW}\) plane.  The color scale shows \(\log_{10}D_{\rm full}\), where \(D_{\rm full}\) is the transfer-function residual norm defined in Appendix~\ref{app:profile_residual}.  The cored-NFW truth has \(M_{\rm vir}=10^7\,M_\odot\), \(z_{\rm L}=0.5\), \(z_{\rm S}=2.0\), \(y_{\rm true}=3\), \(c_{\rm true}=10\), and \(r_c/r_s=0.1\).  The dashed horizontal line marks the true cored-NFW impact parameter, while the star denotes the best-fit NFW template.
    }
    \label{fig:Dfull_heatmap}
\end{figure}

Figure~\ref{fig:Dfull_heatmap} shows the two-dimensional scan of the NFW-template residual in the \(y_{\rm NFW}\)-\(c_{\rm NFW}\) plane.  We plot \(\log_{10}D_{\rm full}\), where \(D_{\rm full}\) is the profile-level residual norm defined in Appendix~\ref{app:profile_residual}.  The logarithmic color scale makes the narrow low-residual valley and the best-fit region visually identifiable.  This plot should be read as a transfer-function degeneracy diagnostic; the detector-weighted mismatch is introduced in Sec.~\ref{sec:detectability}.

The heat map shows that the best-fit region is not centered at the true concentration \(c_{\rm true}=10\).  Instead, the best-fit NFW template occurs at \(y^*\simeq 2.94\) and \(c^*\simeq 8.03\), with a small but nonzero minimum residual.  The impact parameter remains close to the true value \(y_{\rm true}=3\), while the concentration is shifted to a lower value.  This behavior has a clear physical interpretation.  A cored halo has a shallower inner mass distribution than an NFW halo with the same virial mass.  The NFW template can partially mimic this shallower central potential by reducing its concentration.  In other words, the fitting procedure uses a lower-concentration NFW halo to imitate the effect of removing the cusp.

However, the low-residual region is localized rather than extended over the entire parameter plane.  Moving away from the best-fit valley, either by changing \(y_{\rm NFW}\) or by changing \(c_{\rm NFW}\), rapidly increases the residual.  This indicates that the degeneracy between a cored halo and an NFW halo is only partial.  The NFW family can reproduce some broad features of the cored-NFW amplification factor, but it cannot freely absorb the full core-induced spectral distortion.

\subsection{Best-fit residuals}
\label{subsec:bestfit_residuals}

\begin{figure}[t]
    \centering
    \includegraphics[width=\columnwidth]{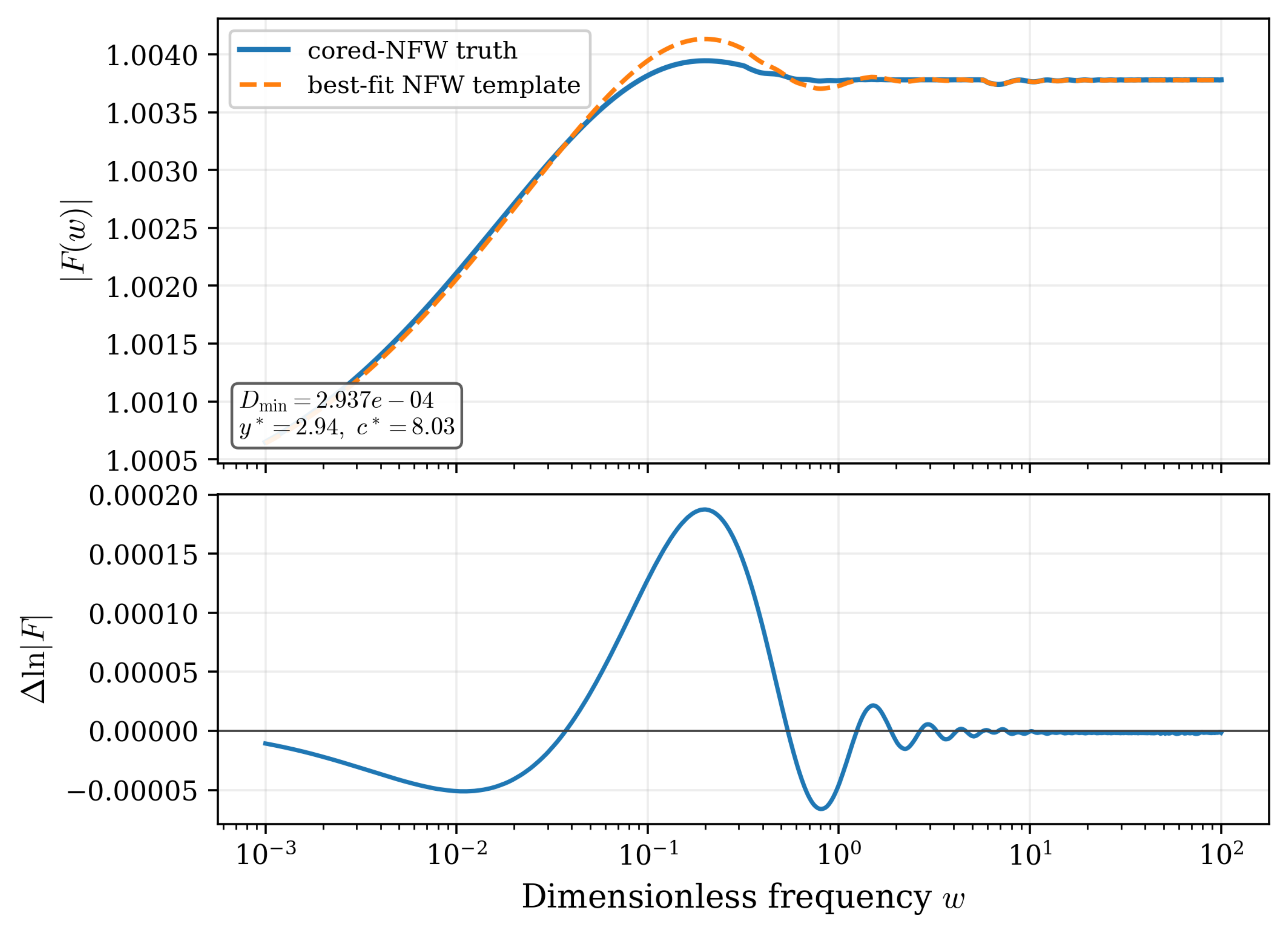}
    \caption{
    Best-fit NFW template compared with the cored-NFW truth.  The upper panel shows \(|F(w)|\) for the cored-NFW truth and the best-fit NFW template.  The lower panel shows the logarithmic amplitude residual \(\Delta\ln|F|\).  Although the best-fit NFW template closely follows the overall amplification curve, a structured residual remains around the diffraction-transition regime.
    }
    \label{fig:bestfit_residuals}
\end{figure}

The best-fit comparison is shown more explicitly in Fig.~\ref{fig:bestfit_residuals}.  The upper panel compares the magnitude of the amplification factor for the cored-NFW truth and the best-fit NFW template.  The lower panel shows the logarithmic amplitude residual,
\begin{equation}
\Delta\ln |F|(w)
=
\ln |F_{\rm NFW}^{\rm best}(w)|
-
\ln |F_{\rm cNFW}(w)| .
\label{eq:log_amp_residual}
\end{equation}
The phase and time-delay alignment does not affect this magnitude residual, but it is included when determining the best-fit parameters.

The upper panel shows that varying \(y_{\rm NFW}\) and \(c_{\rm NFW}\) produces a close match to the overall shape of \(|F(w)|\), including the broad rise toward the first diffraction peak and the approximate high-frequency level.  Thus part of the core effect is degenerate with ordinary NFW lens parameters, especially concentration.

The lower panel, however, reveals a residual that is clearly structured in frequency.  The residual is not a constant offset, nor is it a simple normalization mismatch.  Instead, it changes sign and is concentrated mainly around the diffraction-transition region, \(w\sim 10^{-2}\text{--}1\).  This is precisely the frequency range in which the amplification factor is most sensitive to the detailed shape of the Fermat-potential contours.  At higher frequencies the residual becomes smaller and oscillatory, consistent with the approach toward the geometric-optics regime.

This result is central to the interpretation of the template scan.  Even after allowing the NFW template to adjust its impact parameter and concentration, and even after minimizing over an overall time shift and constant phase, the cored-NFW truth leaves a residual spectral structure.  Therefore, the cored-halo signature is not fully equivalent to a change in NFW parameters.  It contains additional information about the inner density profile.

\subsection{Dependence on core size}
\label{subsec:Dmin_core_size}

\begin{figure}[t]
    \centering
    \includegraphics[width=\columnwidth]{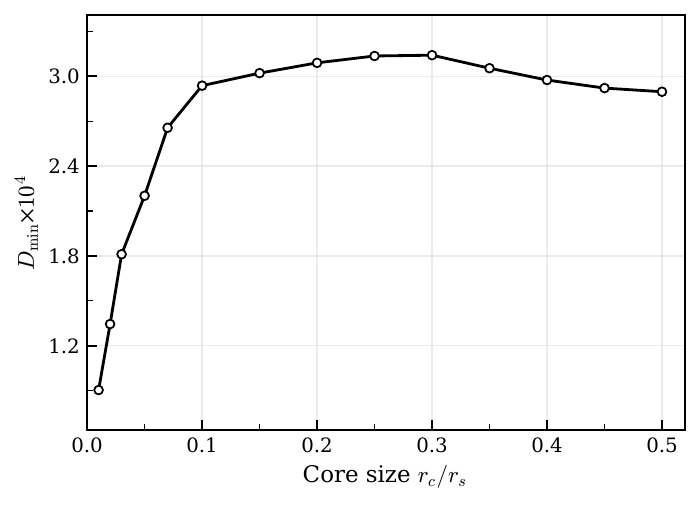}
    \caption{
    Minimum NFW-template residual \(D_{\min}\) as a function of the cored-NFW core size \(r_c/r_s\).  The fiducial truth has \(M_{\rm vir}=10^7\,M_\odot\), \(z_{\rm L}=0.5\), \(z_{\rm S}=2.0\), \(y_{\rm true}=3\), and \(c_{\rm true}=10\).  The distinguishability first increases with core size, reaches a broad maximum around \(r_c/r_s\simeq 0.25\)--\(0.3\), and then decreases slowly.
    }
    \label{fig:Dmin_coretrend}
\end{figure}

We now examine how the residual after NFW-template fitting changes with core size.  We sample \(r_c/r_s=0.01,0.02,0.03,0.05,0.07,0.1\) and then \(0.15\le r_c/r_s\le0.5\) in steps of 0.05, resolving both the NFW recovery regime and the intermediate-to-large-core regime.  For each core size, we repeat the NFW-template scan over \(y_{\rm NFW}\) and \(c_{\rm NFW}\).  The result is shown in Fig.~\ref{fig:Dmin_coretrend}.

The trend is non-monotonic.  For very small cores, the cored-NFW profile is close to NFW, and the best-fit NFW template can reproduce the amplification factor well.  Consequently, the best-fit residual is small.  As \(r_c/r_s\) increases, the inner cusp is progressively flattened.  This produces a larger change in the convergence, lensing potential, and Fermat-potential contours, which in turn enhances the difference in the full complex \(F(w)\).  In this regime, the residual rises rapidly.

The distinguishability reaches a broad maximum around \(r_c/r_s\simeq 0.25\text{--}0.3\).  At this point the core is large enough to generate a substantial wave-optics distortion, but the lensing perturbation is still strong enough for the difference to appear clearly in the frequency-domain response.  This intermediate-core regime is therefore the most difficult for the NFW template family to mimic.

For even larger cores, the residual decreases slowly.  This does not mean that the density profile becomes more NFW-like.  Rather, two effects compete.  First, making the core larger further suppresses the central convergence and makes the inner potential shallower, reducing the strength of the central diffraction feature.  Second, a lower-concentration NFW template can increasingly mimic the broad effect of a shallower inner potential, even though it cannot reproduce the detailed cored profile point by point.  The residual therefore remains nonzero, but the integrated residual over the frequency band decreases mildly.

The non-monotonic behavior of the best-fit residual is thus physically meaningful.  It reflects a competition between two effects.  On the one hand, increasing the core size makes the density profile and lensing potential more different from those of an NFW halo.  On the other hand, once the core becomes too large, the central region becomes less efficient at producing sharp diffraction features, and the resulting smoother response can be partially mimicked by a lower-concentration NFW template.  This also emphasizes that the distinguishability of a core is not determined solely by its size.  It depends on how the core reshapes the Fermat-potential contours and how much of that reshaping can be absorbed by ordinary NFW parameters.

\subsection{NFW recovery test}
\label{subsec:nfw_recovery_test}

We perform an NFW recovery test to verify that the cored-NFW calculation continuously approaches the NFW result in the limit \(r_c/r_s\rightarrow0\).  This test is distinct from the NFW-template fitting analysis above.  Here no NFW-template parameters are varied: the NFW and cored-NFW halos are compared at the same \(M_{\rm vir}\), \(z_{\rm L}\), \(z_{\rm S}\), \(c_{\rm vir}\), and impact parameter \(y\).  The purpose is therefore to check the consistency of the cored-NFW mass normalization and of the full wave-optics calculation pipeline.

We quantify the recovery using the relative difference between the full complex amplification factors.  We define the recovery residual as
\begin{align}
\Delta F_{\rm rec}(w;r_c/r_s)
&\equiv
F_{\rm cNFW}(w;r_c/r_s)
\nonumber\\
&\quad
-
e^{i(w\Delta\tau+\varphi_0)}
F_{\rm NFW}(w),
\label{eq:deltaF_rec}
\end{align}
where \(\Delta\tau\) and \(\varphi_0\) account for an overall time shift and a constant phase offset.  The recovery error is then defined as
\begin{equation}
\epsilon_{\rm rec}(r_c/r_s)
=
\min_{\Delta\tau,\varphi_0}
\left[
\frac{
\int_{w_{\min}}^{w_{\max}}
|\Delta F_{\rm rec}(w;r_c/r_s)|^2\,d\ln w
}{
\int_{w_{\min}}^{w_{\max}}
|F_{\rm NFW}(w)|^2\,d\ln w
}
\right]^{1/2}.
\label{eq:epsilon_rec}
\end{equation}
This quantity measures the fractional difference between the cored-NFW and NFW amplification factors after removing only the two nuisance degrees of freedom associated with an overall time and phase shift.  A successful recovery requires \(\epsilon_{\rm rec}\rightarrow0\) as \(r_c/r_s\rightarrow0\).

We perform this test for the fiducial parameters \(M_{\rm vir}=10^7\,M_\odot\), \(z_{\rm L}=0.5\), \(z_{\rm S}=2.0\), \(c_{\rm vir}=10\), and \(y=3\), using the same frequency interval \(10^{-3}\le w\le10^2\) as in the template-fitting analysis.  The results are listed in Table~\ref{tab:nfw_recovery}.

\begin{table}[t]
\caption{
Recovery error for the cored-NFW model in the \(r_c/r_s\rightarrow0\) limit.
}
\label{tab:nfw_recovery}
\begin{ruledtabular}
\begin{tabular}{cc}
\(r_c/r_s\) & \(\epsilon_{\rm rec}\) \\
\hline
0.10  & \(6.00\times10^{-4}\) \\
0.03  & \(2.25\times10^{-4}\) \\
0.01  & \(8.26\times10^{-5}\) \\
0.003 & \(2.61\times10^{-5}\) \\
0.001 & \(8.87\times10^{-6}\) \\
\end{tabular}
\end{ruledtabular}
\end{table}

The recovery error decreases monotonically as \(r_c/r_s\) is reduced.  In particular, when the core size is decreased from \(r_c/r_s=0.1\) to \(r_c/r_s=10^{-3}\), \(\epsilon_{\rm rec}\) drops from \(6.00\times10^{-4}\) to \(8.87\times10^{-6}\).  This confirms that the cored-NFW result approaches the NFW result continuously at the level of the full complex frequency-domain response.  The recovery test therefore provides an end-to-end numerical consistency check of the fixed-mass cored-NFW construction and the wave-optics pipeline used throughout this work.

\section{Detector-weighted relevance for a fiducial LISA source}
\label{sec:detectability}

The previous sections show that finite cores modify the transfer function.  We now ask whether this modification can matter for a representative LISA MBHB waveform.  This section is only a scale estimate.  It is not a Bayesian parameter-estimation study and it is not an event-rate forecast.  We compute the mismatch between a cored-NFW lensed waveform and an unlensed waveform to identify where the lensing distortion itself becomes visible.  A full NFW-versus-cored-NFW comparison would use the same inner product between the best-fit NFW-lensed and cored-NFW-lensed waveforms while also varying source parameters and the full time-dependent LISA response.

For a cored-NFW lens we write
\begin{equation}
\begin{aligned}
\tilde h_{\rm L}(f;M_{\rm vir},y,q)&=F[w(f;M_{\rm vir},z_{\rm L}),y,q]\tilde h_0(f),\\
w(f;M_{\rm vir},z_{\rm L})&=8\pi G(1+z_{\rm L})M_{\rm vir}f/c^3 .
\end{aligned}
\label{eq:lensed_waveform_detectability}
\end{equation}
Here $\tilde h_0(f)$ is the unlensed waveform, $\tilde h_{\rm L}(f)$ is the lensed waveform, and $q=r_c/r_s$ is the cored-NFW core size.  We use the standard mismatch of Eq.~\eqref{eq:standard_mismatch} with $h_1=h_{\rm L}$ and $h_2=h_0$ and denote the result by ${\cal M}_{\rm L0}$.  The inner product uses a sky-averaged LISA-like noise curve including the four-year Galactic binary confusion foreground~\cite{2019CQGra..36j5011R}.  We then form
\begin{equation}
S(M_{\rm vir},y)\equiv {\rm SNR}_0^2 {\cal M}_{\rm L0}(M_{\rm vir},y),\qquad
{\rm SNR}_0^2=\langle h_0|h_0\rangle,
\label{eq:S_detectability}
\end{equation}
and take $S>1$ as a practical threshold for a potentially measurable lensing-induced distortion, following the usual $\rho^2{\cal M}\gtrsim1$ criterion used in waveform comparisons~\cite{2025PhRvD.111b4068B}.  The quantity ${\rm SNR}_0$ is the optimal signal-to-noise ratio of the unlensed waveform under the adopted noise curve.

The scan uses $q=r_c/r_s=0.3$, $c_{\rm true}=10$, $z_{\rm L}=0.5$, and $z_{\rm S}=2.0$.  The source is a fiducial equal-mass MBHB with total mass $M_{\rm BBH}=10^6\,M_\odot$, modeled with IMRPhenomXHM~\cite{2020PhRvD.102f4002G}.  We evaluate the mismatch over $10^{-5}\,{\rm Hz}\le f\le0.1\,{\rm Hz}$, set the inclination to $\iota=0$, and use a response factor equal to unity.  These assumptions define an optimistic high-SNR scale estimate.

\begin{figure}[t]
    \centering
    \includegraphics[width=\columnwidth]{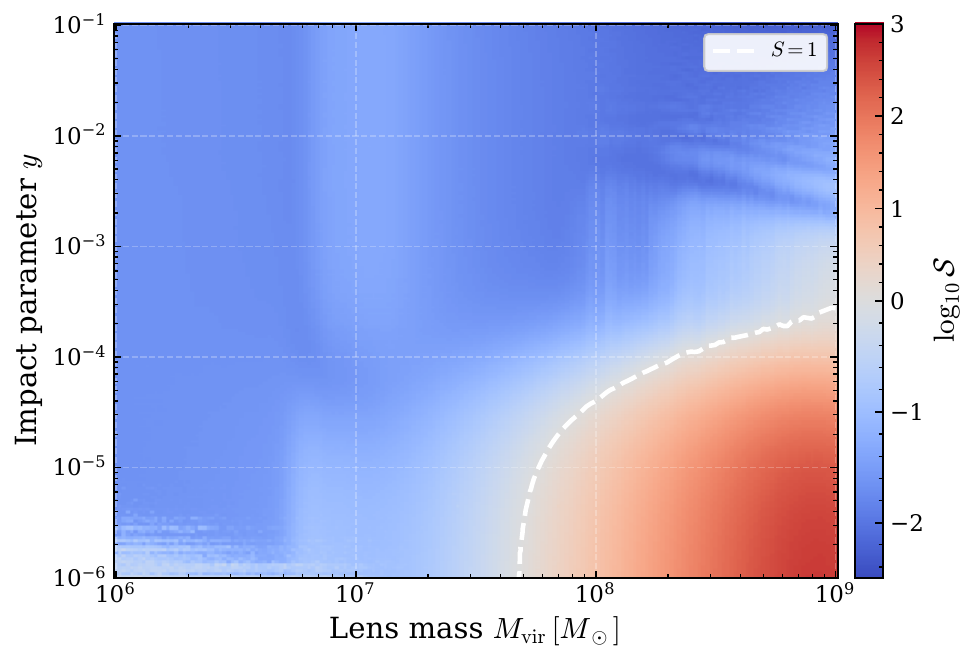}
    \caption{
    Detector-weighted relevance of the cored-NFW lensing distortion for the fiducial LISA MBHB source.  The color scale shows $\log_{10}S$, where $S\equiv{\rm SNR}_0^2{\cal M}_{\rm L0}$.  The dashed white curve marks $S=1$.  The scan uses $r_c/r_s=0.3$, $c_{\rm true}=10$, $z_{\rm L}=0.5$, $z_{\rm S}=2.0$, and $M_{\rm BBH}=10^6\,M_\odot$.  The overlap is maximized over a coalescence-time shift and a constant phase offset.
    }
    \label{fig:detectability_map}
\end{figure}

Figure~\ref{fig:detectability_map} shows that the detectable region is restricted to relatively massive lenses and very small impact parameters.  For the adopted source, halos below a few $\times10^7\,M_\odot$ remain below $S=1$ throughout the scanned range.  At larger lens masses the allowed impact parameter grows, but it remains small: the effect is a favorable-alignment signal, not a generic distortion expected for every low-mass halo crossing the line of sight.

This result also clarifies the relation between the isolated-halo and strong-lens interpretations.  For an isolated halo, the small-$y$ requirement corresponds to a small geometrical cross section.  In a strongly lensed GW system, a subhalo near a highly magnified macro-image could be more favorable because the macro-lens boosts the image signal-to-noise ratio and fixes a region of high lensing sensitivity.  However, that regime requires adding the macro convergence, shear, macro-image parity, and a realistic subhalo population.  These effects are outside the present paper.  The present detectability map should therefore be read as the observational scale of the generic finite-core transfer function, not as a final prediction for the number of detectable events.

We restrict the scan to $M_{\rm vir}\le10^9\,M_\odot$ because, for the fiducial LISA-band source, still larger lenses enter a mostly geometric-optics-like regime over much of the frequency interval carrying the signal-to-noise ratio.  The mass range shown therefore focuses on the diffractive transition where the inner density profile can leave frequency-dependent information.

\section{Discussion and conclusions}
\label{sec:conclusions}

We have studied a deliberately focused problem: what part of a wave-optics dark-matter lensing signal is the generic consequence of replacing an NFW cusp by a finite-density core?  This is a prerequisite for interpreting more model-specific signatures such as FDM solitons or SIDM isothermal cores.  We used a one-parameter cored-NFW profile as a finite-core baseline, normalized it at fixed virial mass, and compared it with the corresponding cuspy NFW halo.

The main physical effect is simple.  A finite core lowers the central convergence and makes the inner lensing potential shallower.  This changes the Fermat-potential contours sampled by the gravitational wave and therefore broadens the time-domain response $I(\tau)$ and changes the diffraction peak and oscillatory structure of the complex amplification factor $F(w)$.  The SIDM-inspired IC-NFW comparison produces the same qualitative behavior, supporting the interpretation that the effect is a generic finite-core response rather than an artifact of one analytic parameterization.

The main degeneracy is also clear.  An NFW template can imitate part of a cored halo by lowering its concentration, because both changes reduce the central projected mass.  However, the degeneracy is incomplete.  After varying the NFW impact parameter and concentration, and after minimizing over an overall time shift and constant phase, structured residuals remain in the complex frequency-domain response.  The profile-level residual is non-monotonic in core size: it is small as $r_c/r_s\rightarrow0$, peaks around $r_c/r_s\simeq0.25$--$0.3$, and decreases mildly for larger cores as the central diffraction feature becomes smoother and easier to mimic with a lower-concentration NFW profile.  The explicit recovery test confirms that the cored-NFW calculation approaches NFW smoothly in the small-core limit.

The detector-weighted estimate places the effect in a realistic scale.  For the fiducial LISA MBHB source and the optimistic response assumptions used here, cored-NFW lensing distortions satisfy $S={\rm SNR}_0^2{\cal M}_{\rm L0}>1$ mainly for $M_{\rm vir}\gtrsim 10^7\,M_\odot$ and very small impact parameters.  Thus the effect is not a generic large-cross-section signal from arbitrary low-mass halos.  It is most relevant for favorable near-aligned isolated lenses or for subhalos perturbing highly magnified macro-images in strongly lensed systems.  The latter possibility is particularly interesting but is not fully modeled in this paper.

Several limitations should therefore be kept explicit.  We considered an isolated, axially symmetric local perturber and did not include external convergence, external shear, macro-image parity, triaxiality, baryons, halo-to-halo scatter, line-of-sight structure, or a cosmological subhalo population.  The detector calculation used a single fiducial MBHB source, an analytic sky-averaged LISA-like noise curve, and did not perform a full source-plus-lens parameter estimation.  Future work should embed cored and cuspy subhalos in realistic galaxy-scale strong lenses, include macro magnification and parity, propagate the full time-dependent LISA response, and compare NFW, cored-NFW, FDM, and SIDM profiles with the waveform mismatch and Bayesian evidence.  A natural extension is to build simulated catalogs and then use Bayesian sampling or machine-learning classification to separate profile classes in a population-level analysis.

Within these limitations, the paper establishes a clean finite-core transfer-function baseline.  Finite central cores leave coherent wave-optics imprints that are partly, but not completely, degenerate with ordinary NFW lens parameters.  This baseline is necessary before claiming a future diffractive GW lensing signal as evidence for any particular dark-matter microphysics.

\appendix

\section{Transfer-function residual used in the NFW-template scan}
\label{app:profile_residual}

The detector-level comparison in Sec.~\ref{sec:detectability} uses the standard mismatch of Eq.~\eqref{eq:standard_mismatch}.  For the profile-level figures, where no detector noise curve or source waveform is specified, we used a simpler complex residual norm to locate the NFW degeneracy valley.  For a cored-NFW truth and an NFW template, we write
\begin{align}
\Delta F_{\rm full}(w)
&=F_{\rm NFW}(w;y_{\rm NFW},c_{\rm NFW})
-e^{i(w\Delta\tau+\varphi_0)}F_{\rm cNFW}(w),
\end{align}
and minimize over the nuisance time-delay and phase offsets.  The plotted residual is
\begin{equation}
D_{\rm full}=\min_{\Delta\tau,\varphi_0}
\left[\int_{w_{\min}}^{w_{\max}}|\Delta F_{\rm full}(w)|^2d\ln w\right]^{1/2},
\end{equation}
with $10^{-3}\le w\le10^2$.  The minimum over the NFW template parameters is denoted by $D_{\min}$.  This quantity is useful for visualizing the response-level degeneracy, while the observational statistic for real data should be the detector-weighted mismatch or a full likelihood.

\section{Numerical definitions used in the lensing pipeline}
\label{app:numerical_definitions}

This appendix collects technical definitions that are used in the numerical calculation but are not central to the main physical argument.  We assume a spatially flat $\Lambda$CDM cosmology with $H_0=70~{\rm km~s^{-1}~Mpc^{-1}}$, $\Omega_{\rm m}=0.3$, and $\Omega_\Lambda=0.7$.  With $a_{\rm L}=1/(1+z_{\rm L})$, and using the angular-diameter distances defined in Sec.~\ref{subsec:astrophysical_setup}, we define
\begin{equation}
d_{\rm eff}=a_{\rm L}\frac{D_{\rm L}D_{\rm LS}}{D_{\rm S}},
\qquad
\Sigma_{\rm cr}=\frac{a_{\rm L}}{4\pi G d_{\rm eff}} .
\label{eq:deff_sigma_app}
\end{equation}
The expression for $\Sigma_{\rm cr}$ follows the geometric-unit convention used in the main text; a factor of $c^2$ should be inserted in the numerator when SI units are restored.  The projected surface density of a spherically symmetric halo is
\begin{equation}
\Sigma(\xi)=\int_{-\infty}^{+\infty}\rho\!\left(\sqrt{\xi^2+Z^2}\right)dZ,
\qquad
\kappa(x)=\frac{\Sigma(\xi_0 x)}{\Sigma_{\rm cr}} ,
\label{eq:surface_density_app}
\end{equation}
where $Z$ is the physical line-of-sight coordinate, $\xi_0^2=4G M_{\rm Lz}d_{\rm eff}$ is the lens-plane length scale, and $x=\xi/\xi_0$ is the dimensionless radius used internally by the wave-optics integral.  For numerical projection of the cored profiles we use $Z=\xi\sinh t$, giving
\begin{equation}
\Sigma(\xi)=2\xi\int_0^\infty \rho(\xi\cosh t)\cosh t\,dt .
\label{eq:projection_t}
\end{equation}
For an axially symmetric lens we compute the projected mass $m(x)=2\int_0^x\kappa(x')x'dx'$ and then $d\psi/dx=m(x)/x$, up to an irrelevant additive constant in $\psi$.

\section{IC-NFW matching equations}
\label{app:icnfw_details}

For the SIDM-inspired IC-NFW cross-check, the matching radius is $r_1=s r_s$.  The inner density is written as
\begin{equation}
\rho_{\rm iso}(r)=\rho_{\rm iso,0}\exp[-h(Q)],
\qquad Q=\frac{r}{r_0},
\label{eq:iso_density_app}
\end{equation}
where the self-gravitating isothermal functions satisfy
\begin{equation}
\frac{dh}{dQ}=\frac{m(Q)}{Q^2},
\qquad
\frac{dm}{dQ}=Q^2\exp[-h(Q)] .
\label{eq:isothermal_eqs_app}
\end{equation}
Continuity of density and enclosed mass at $r_1$ leads to
\begin{equation}
\frac{m(Q_1)e^{h(Q_1)}}{Q_1^3}
=
A(s)\frac{(1+s)^2}{s^2},
\
A(s)=\ln(1+s)-\frac{s}{1+s},
\label{eq:Q1_matching_app}
\end{equation}
with $Q_1=r_1/r_0$.  Once $Q_1$ is found, the isothermal parameters are
\begin{equation}
r_0=\frac{s r_s}{Q_1},
\qquad
\rho_{\rm iso,0}=\frac{\rho_s e^{h(Q_1)}}{s(1+s)^2} .
\label{eq:iso_params_app}
\end{equation}
These equations specify the IC-NFW profile used only for the qualitative cross-check in Fig.~\ref{fig:sidm_comparison}.

\section*{Acknowledgments}
This work was supported by National Key R$\&$D Program of China (No. 2024YFC2207400); the National Natural Science Foundation of China under Grants No. 12475051, and No. 12035005; the Innovative Research Group of Hunan Province under Grant No. 2024JJ1006; the Science and Technology Innovation Program of Hunan Province under Grant No. 2024RC1050; and the Natural Science Fund of Hunan Province under Grant No. 2026JJ20019.
\bibliography{references}

\end{document}